\documentclass[]{article}

\usepackage[margin=0.75in]{geometry}
\usepackage[capposition=top]{floatrow}
\usepackage{xcolor}
\usepackage{amsmath}
\usepackage{amsfonts}
\usepackage{amsthm}
\usepackage{graphicx}
\usepackage{float}
\usepackage{hyperref}
\usepackage{setspace}
\usepackage{multicol}
\usepackage{lipsum}
\usepackage{algorithm}
\usepackage{algpseudocode}

\theoremstyle{definition}

\title{Constructing g-computation estimators: two case studies in selection bias}
\author{Paul N Zivich\textsuperscript{1}, Haidong Lu\textsuperscript{2}}
\date{%
	\textsuperscript{1}Department of Epidemiology, Gillings School of Global Public Health, University of North Carolina at Chapel Hill, Chapel Hill, NC\\%
	\textsuperscript{1}Department of Internal Medicine, Yale School of Medicine, Yale University, New Haven, CT\\%
	~\\
	\today
}

\begin{document}
	
\maketitle

\begin{abstract}
	G-computation is a useful estimation method that can be adapted to address various biases in epidemiology. However, these adaptations may not be obvious for some complex causal structures. This challenge is an example of the much wider issue of translating a causal diagram into a novel estimation strategy. To highlight these challenges, we consider two recent cases from the selection bias literature: treatment-induced selection and co-occurrence of biases that lack a joint adjustment set. For each case study, we show how g-computation can be adapted, described how to implement that adaptation, show some general statistical properties, and illustrate the estimator using simulation. To simplify both the theoretical study and practical application of our estimators, we express the proposed g-computation estimators as stacked estimating equations. These examples illustrate how epidemiologists can translate identification results into a g-computation estimator and study the theoretical and finite-sample properties of a novel estimator.
\end{abstract}

\section{Introduction}

A practical challenge to empirical epidemiologic research is translating from a causal diagram identification result to an estimator. While the epidemiologic and statistical literature is filled with novel estimators, these approaches are often tailored to specific scenarios. For example, an estimator may not deal with multiple sources of bias (e.g., confounding, selection bias, missing data), or assume there is an overall sufficient adjustment set for all the sources of systematic errors \cite{Zivich2022On}. The gap between drawing the causal diagram and developing an estimation strategy may dissuade some researchers from considering these novel estimators, even when they are thought to produce more reliable estimates relative to traditional regression methods.

Here, we consider methods to address selection bias as a case study of translating from identification results to estimation. Recent work has focused on clarifying the definition of selection bias \cite{Lu2022Toward,Lu2023Selection, Lu2024Evolution}, and provided graphical rules for addressing selection bias \cite{Kenah2023Potential,Mathur2024Simple}. However, these studies do not immediately provide corresponding estimators to recover treatment effects from selection bias. When faced with selection bias, epidemiologists may default to inverse probability weighting (IPW) estimators, as they allow for one to independently model different sources of bias (e.g., inverse probability of treatment weights for confounding, inverse probability of censoring or sampling weights for different types of selection bias) \cite{Robins2000Marginal}. While IPW provides a fairly general recipe, there are can be challenges to its application \cite{Zivich2025Estimating}. First, an epidemiologist must correctly specify each IPW model, which can be difficult as it requires sufficient but, oftentimes, distinct background knowledge for each source of bias. Second, the order in which IPW models are constructed can differ depending on the context \cite{Ross2021Reflection}. Third, IPW estimators are generally less efficient (i.e., wider confidence intervals) relative to competing estimators \cite{Daniel2018Double}. These issues can be avoided by adopting outcome-model-based approaches (e.g., g-computation \cite{Snowden2011Implementation}), which only require correct specification of a model for the outcome process (avoiding separate models for each source of bias and concern over the ordering of estimation of different weights) and is generally more efficient \cite{Zivich2025Estimating, Chatton2020G}. However, g-computation may need to be adapted for certain selection bias mechanisms or under the joint occurrences of biases (e.g., confounding and selection bias) that lack a sufficient adjustment superset \cite{Zivich2022On,Breskin2018Practical}. Therefore, adaptations of the standard g-computation algorithm are needed for causal structures under these scenarios.

Using two recent examples in the selection bias literature as case studies, we illustrate how epidemiologists can translate identification results into novel g-computation estimators. For completeness, we rederive the motivating identification results, propose estimators based on those identification results, prove some general properties of our estimators, and illustrate their application using Monte Carlo simulation experiments. For the first case study, we consider treatment-induced selection bias. In the second case study, we study the joint occurrence of confounding and selection bias where there is not an overlapping adjustment set.  While estimators are developed for these specific cases, the central ideas of this work can be used to build g-computation estimators in other contexts.

\section{Estimating Equations}

To develop our estimators, we rely on M-estimation theory. For more in-depth introductions, see the following references \cite{Stefanski2002Calculus,Boos2013M,Ross2024M,Desmond2014Estimating,Jesus2011Estimating}. An M- or Z-estimator, $\hat{\theta}$, is defined as the solution to 
\begin{equation*}
	\frac{1}{n} \sum_{i=1}^{n} g(Z_i; \hat{\theta}) = 0
\end{equation*}
where $\sum_{i=1}^{n} g(Z_i; \hat{\theta})$ is referred to as the estimating equation, $g(Z_i; \theta)$ is called an estimating function, $Z_i$ is the observed data for individual $i$, and $\theta$ are the parameters. Colloquially, an M-estimator is simply the point at which the estimating equation is equal to zero (i.e., $\hat{\theta}$ corresponds to the `root' of the estimating equation). Note that $g$ and $\theta$ are both $k$-dimensional (i.e., if there are three parameters, $\theta = (\theta_1, \theta_2, \theta_3)$, $g$ is a function that returns a vector of three scalars and the solution is where the estimating equation is equal to $(0,0,0)^T$).

M-estimators offer four benefits for us. First, M-estimators provide a general recipe for estimating the variance of parameters via the so-called empirical sandwich variance estimator \cite{Boos2013M, Raymond2006Measurement}. By jointly solving a stack of estimating equations, the empirical sandwich variance estimator appropriately incorporates the dependence of parameters on other parameters. Essentially, we can estimate the variance of the parameter of interest, while correctly incorporating the uncertainty in estimation of the outcome model for g-computation. This feature allows us to avoid more computationally intensive methods for variance estimation, such as the bootstrap \cite{Efron1979Bootstrap, Kulesa2015Sampling}. However, any variable selection procedure not based on domain knowledge needs to be \textit{part} of the stacked estimating equations for valid variance esitmation. Note that the empirical sandwich variance estimator can also be used with IPW estimators \cite{Reifeis2022On}. Second, M-estimation theory can simplify consistency and asymptotic normality proofs for estimators. Hence, we can more easily prove certain statistical guarantees for our estimator as the sample size approaches infinity. Briefly, if the estimating equations are shown to be unbiased at the true $\theta$, it suffices to show that the corresponding estimator is consistent and asymptotically normal following some additional regularity conditions \cite{Boos2013M}. This underlying statistical theory is what justifies use of the empirical sandwich variance estimator. As we propose novel estimators, we use this feature to justify our estimators. Third, many of the estimators that epidemiologists routinely use can be readily expressed with estimating equations. For example, both the arithmetic mean and generalized linear model can be expressed as estimating equations \cite{Zivich2025Estimating, Boos2013M}. Therefore, the underlying pieces of the estimators proposed in this paper are connected to approaches already used by epidemiologists. Finally, there is free and open-source software for the automated computation of M-estimators \cite{Zivich2022Delicatessen, Saul2020Calculus}, simplifying their application.

\section{Standard G-computation}

\begin{figure}[H]
	\centering
	\caption{Single world intervention graph for confounding and selection bias with a sufficient adjustment superset}
	\includegraphics[width=0.5\linewidth]{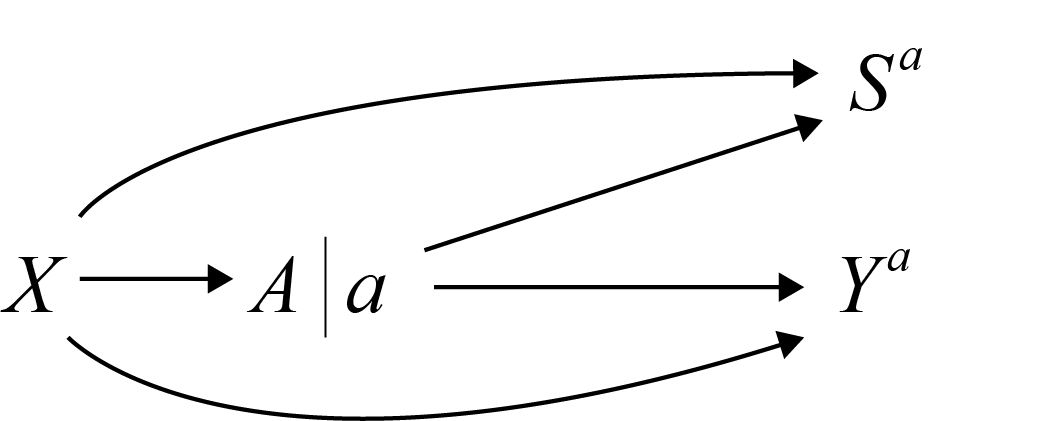}
	\label{Figure1}
\end{figure}

Before the case studies, we review a standard g-computation implementation for the average causal effect with selection bias and how it can be implemented as an M-estimator. Let $Y$ denote the outcome, $A$ denote the treatment, $S$ denote whether a person completed follow-up (i.e., $Y$ is only observed for those with $S=1$), and $X$ be a set of baseline covariates. Following the Single-World Intervention Graph (SWIG) in Figure \ref{Figure1} \cite{Breskin2018Practical}, $X$ is a sufficient adjustment set for both confounding and selection bias due to differential loss to follow-up. The parameter of interest is $\psi = \mu^1 - \mu^0$, where $\mu^a = E[Y^a]$, $Y^a$ is the potential outcome under treatment $a$, and $E[\cdot]$ is the expected value function. $\psi$ can be re-expressed in terms of the observed data as 
\begin{equation}
	\mu^a = E \left[ E(Y \mid A=a, X, S=1) \right]
	\label{ID-R1}
\end{equation}
following conditional exchangeability with positivity for both confounding and selection bias, and causal consistency (see Appendix 1 for details) \cite{Hernan2006Estimating, Cole2009consistency, Zivich2022Positivity}.

Using the expression in \ref{ID-R1} and the algorithm described by Snowden et al. adapted for missing outcome \cite{Snowden2011Implementation}, we can estimate $\psi$. First, we fit an outcome model for $Y$ given $A,X$ among those who completed follow-up (i.e., $S=1$). Using that model, we then generate predict values of $Y$ under $a$ for all observations in the study sample (both $S=1$ and $S=0$), which we refer to as pseudo-outcomes and denote as $\hat{Y}^a$. Then we take the mean of the pseudo-outcomes, $\hat{\mu}^a = n^{-1} \sum_{i=1}^n \hat{Y}^a_i$, where $n$ is the size of the entire study sample. By repeating this process for $a:=1$ and $a:=0$, we can then estimate $\psi$ by taking the difference.

Now consider the estimating functions that correspond to these discrete steps,
\begin{equation*}
	g_1(Z_i; \theta) = 
	\begin{bmatrix}
		S_i \left[ Y_i - m(\mathbb{X}_i; \beta)  \right] \mathbb{X}_i^T \\
		\hat{Y}_i^1 - \mu^1 \\
		\hat{Y}_i^0 - \mu^0 \\
		(\mu^1 - \mu^0) - \psi
	\end{bmatrix}
\end{equation*}
where $\theta = (\beta, \mu^1, \mu^0, \psi)$. Here, the first estimating function is for the outcome model restricted to those who completed follow-up, the second and third are for the means of the pseudo-outcomes, and the final estimating function is for the average causal effect. Note that in the first estimating function, $m(\mathbb{X}) = E[Y \mid A,X; \beta]$ with the design matrix $\mathbb{X}$ (e.g., $\mathbb{X} = (1, A, X)$ with $\beta=(\beta_0, \beta_1, \beta_2)$). In the case of a continuous outcome, $m$ may be linear regression (i.e., $m(\mathbb{X}; \beta) = \mathbb{X} \beta^T$). If $Y$ is binary, we might instead use a logistic regression model (i.e., $m(\mathbb{X}; \beta) = \text{expit}(\mathbb{X} \beta^T)$ where $\text{expit}$ is the inverse logit function). While these estimating functions may seem to have magically appeared, the first estimating function is simply the score function for the corresponding regression model \cite{Boos2013M}. One could derive this estimation equation by taking the derivative of the log-likelihood for the chosen regression model \cite{Ross2024M}. As the score functions for many commonly used regression models are well-known results, we simply use those results directly. A similar case holds for the estimating equation for the mean, where $\hat{\mu}^1 = n^{-1} \sum_{i=1}^n \hat{Y}^1$ can be rewritten following some algebra as $\sum_{i=1}^n \left( \hat{Y}^1 - \hat{\mu}^1 \right) = 0$. A similar process applies for the third function. The final estimating function does not depend on the data and follows immediately from $\hat{\psi} = \hat{\mu}^1 - \hat{\mu}^0$. For a more detailed step-by-step derivation of estimating functions for g-computation see the following reference \cite{Ross2024M}.

Having determined the estimating functions, we can now show that our estimator is consistent and asymptotically normal. To demonstrate this, M-estimation theory (under some additional regularity conditions) requires us to only show that each of the corresponding estimating equations is unbiased at the true parameter value \cite{Boos2013M, Stefanski2002Calculus}. So, we need to show $E[g_1(Z; \theta)] = 0$ holds for the true value of $\theta$. Here, we provide a sketch of an argument with a more formal proof in the Appendix. Note that $E\left\{ S_i \left[Y_i - m(\mathbb{X}_i; \beta) \right] \mathbb{X}_i^T \right\} = 0$ follows directly from standard maximum likelihood theory. For the means, unbiasedness follows from the identification assumptions and correct outcome model specification. Finally, the estimating equation for the average casual effect is unbiased as it is simply a transformation of unbiased parameters. Therefore, the g-computation estimator is consistent and asymptotically normal. Note that use of g-computation is still premised on the identification assumptions (e.g., causal consistency, exchangeability, positivity) and that the outcome model is correctly specified.

While the standard g-computation algorithm is valid for some causal structures with selection bias, the previous procedure might not work for other structures. Below we consider two case studies where standard g-computation fails to recover the average causal effect but a modified g-computation algorithm does.

\section{Case Study 1: Treatment-induced selection}

\begin{figure}[H]
	\centering
	\caption{Single world intervention graph for treatment-induced selection bias}
	\includegraphics[width=0.5\linewidth]{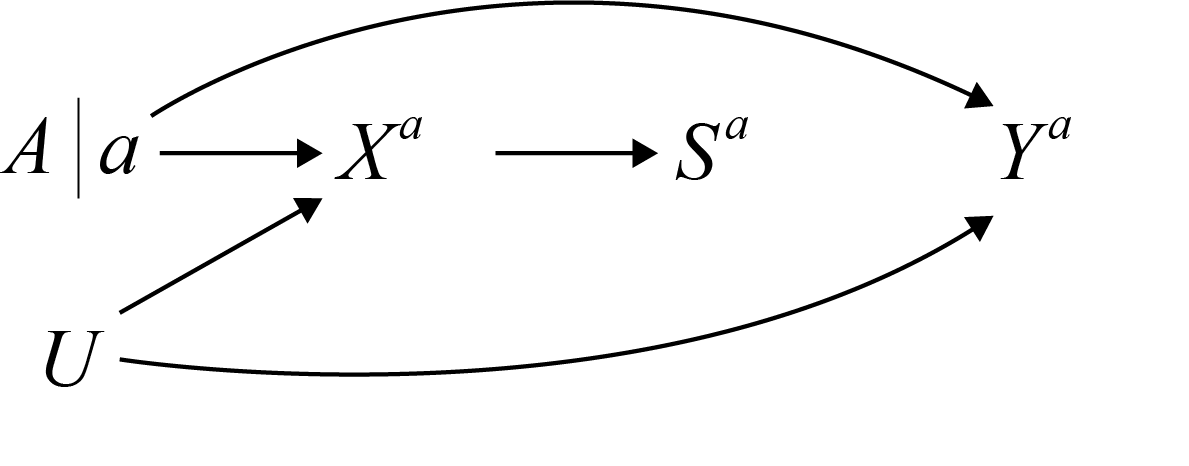}
	\label{Figure2}
\end{figure}

Consider the example described in Breskin et al. and shown in Figure \ref{Figure2} \cite{Breskin2018Practical}. Let $A$ denote vaccination status ($1$: vaccine, $0$: placebo) and $Y$ denote disease at one year ($1$: yes, $0$: no). Here, the vaccine is more likely to result in injection site pain ($X=1$ if pain, $X=0$ otherwise). Additionally, poor overall health ($U$), an unmeasured variable, is related to both the outcome and injection site pain. Finally, $S$ denotes if a person completed follow-up. As such, $Y$ is only observed for those with $S=1$, while $A,X$ are fully observed. The parameter of interest is the average causal effect in the entire trial population, $\psi$. 
Here, $U$ only affects selection through $X$ but standard g-computation that incorporates $X$ will be biased since $X$ is a collider. However, not accounting for $X$ means there is selection bias due to an open backdoor path between $S^a$ and $Y^a$ (type 2 selection bias) \cite{Lu2022Toward}. Therefore, a revised identification strategy is needed. Instead, the interest parameter can be expressed as
\begin{equation}
	\mu^a = E \left[ E(Y \mid A=a, X, S=1) \mid A=a \right]
	\label{ID-R2}
\end{equation}
for $a \in \{0,1\}$ (proof in Appendix 2).This result can how be used to construct a modified g-computation estimator.

\subsection{Building an estimator}

The result in \ref{ID-R2} suggests the following recipe. Starting with the inner expectation, note that $E(Y \mid A,X,S=1)$ can be estimated with a model using only the complete cases. here, $X$ is allowed to be continuous, so we use a parametric model with the parameters $\beta$ (which requires that we assume our model is correctly specified). This model can then be used to generate pseudo-outcomes under $a$ for all observations. The outer expectation indicates that we want to take the mean of the pseudo-outcomes only among those with $A=a$ rather than the \textit{entire} population. Note that this last step is the key difference from the standard g-computation algorithm.

This step-by-step process can be translated into the following set of stacked estimating functions,
\begin{equation*}
	g_2(Z_i; \theta) = 
	\begin{bmatrix}
		S_i \left[ Y_i - m(\mathbb{X}_i; \beta)  \right] \mathbb{X}_i^T \\
		A_i \left(\hat{Y}_i^1 - \mu^1\right) \\
		(1 - A_i) \left(\hat{Y}_i^0 - \mu^0\right) \\
		(\mu^1 - \mu^0) - \psi
	\end{bmatrix}
\end{equation*}
where $\theta = (\beta, \mu^1, \mu^0, \psi)$. This estimating function shares similarities to $g_1$. Like $g_1$, the outcome model is restricted to those who completed follow-up (i.e., $S=1$). However, the estimating functions for the mean are now conditional on values of $A$. These estimating function can be found by re-expressing the formula for the conditional mean as an equation equal to zero. The final estimating function is the average causal effect.

Having determine the estimating equations, we now argue that our proposed estimator is consistent and asymptotically normal with a more formal proof provided in Appendix 2. Note that the outcome model is unbiased again following directly from standard maximum likelihood theory. Similarly, the unbiasedness of the conditional means follows from the identification result in \ref{ID-R2} and correct outcome model specification. Finally, the estimating equation for $\psi$ is unbiased by the definitions of the corresponding parameters.

\subsection{Simulation}

To verify our mathematical derivation and illustrate the performance of the proposed estimator, a modified version of the data generating mechanism from Breskin et al. was created (see Appendix 2 for details). In the simulations, we compare taking the difference between the means of the outcomes among complete-cases stratified by treatment arm (na\"ive analysis), standard g-computation with $X$ (as in \ref{ID-R1}), and our modified g-computation algorithm. For reference, we also include results for an IPW estimator (Appendix 2). For evaluation metrics, we considered bias, empirical standard error (ESE), root mean squared error (RMSE), standard error ratio (SER), and 95\% confidence interval (CI) coverage \cite{Morris2019Using}. Bias was defined as the mean of the difference between estimates and the true $\psi$. ESE was defined as the standard deviation of the point estimates. RMSE was defined as the square root of the mean of the squared difference between estimates and the true $\psi$, where smaller values are better. SER was defined as the average of the estimated standard errors divided by the ESE. As SER near one indicates the variance is being estimated appropriately, while values below one indicate the variance is being underestimated. CI coverage was defined as the proportion of 95\% CIs that covered the true $\psi$. Simulations consisted of $n:=1000$ with 5000 repetitions and were conducted using Python 3.9.4 (Beaveron, OR, USA) using the following packages: \texttt{NumPy} \cite{Harris2020Array}, \texttt{SciPy} \cite{Virtanen2020SciPy}, \texttt{pandas} \cite{McKinney2010Data}, and \texttt{delicatessen} \cite{Zivich2022Delicatessen}. We also replicate the results for a single simulated data set in R 4.4.1 (Vienna, Austria).

Results for the simulation study are shown in Table \ref{Table1}. The na\"ive analysis was biased, as expected. For our specific mechanism, standard g-computation resulted in greater bias relative to the na\"ive analysis. Finally, the modified g-computation and IPW estimators had nearly zero bias and the empirical sandwich variance estimator performed well (as indicated by the SER and nominal 95\% CI coverage), as was expected. In terms of the RMSE, the modified g-computation estimator was the best performing estimator.

\begin{table}[H]
	\caption{Simulation results for treatment-induced selection bias}
	\centering		
	\begin{tabular}{lccccc}
		\hline
		& Bias   & ESE &  RMSE   & SER  & Coverage \\ \cline{2-5} 
		Complete-case analysis (na\"ive) & -0.028 & 0.035 & 0.045 & 1.00 & 87\%     \\
		Standard g-computation                          & -0.144 & 0.035 & 0.148 & 1.00 & 2\%      \\
		Modified g-computation                          & -0.001 & 0.036 & 0.036 & 1.01 & 95\%     \\
		Inverse probability weighting                   & -0.001 & 0.039 & 0.039 & 1.00 & 95\%     \\ \hline
	\end{tabular}
	\floatfoot{ESE: empirical standard error, RMSE: root mean squared error, SER: standard error ratio, CI: confidence interval. \\
	Simulation results were based on a sample size of 1000 observations repeated for 5000 iterations.}
	\label{Table1}
\end{table}

\section{Case Study 2: Confounding and selection bias}

\begin{figure}[H]
	\centering
	\caption{Single world intervention graph for treatment-induced selection bias}
	\includegraphics[width=0.5\linewidth]{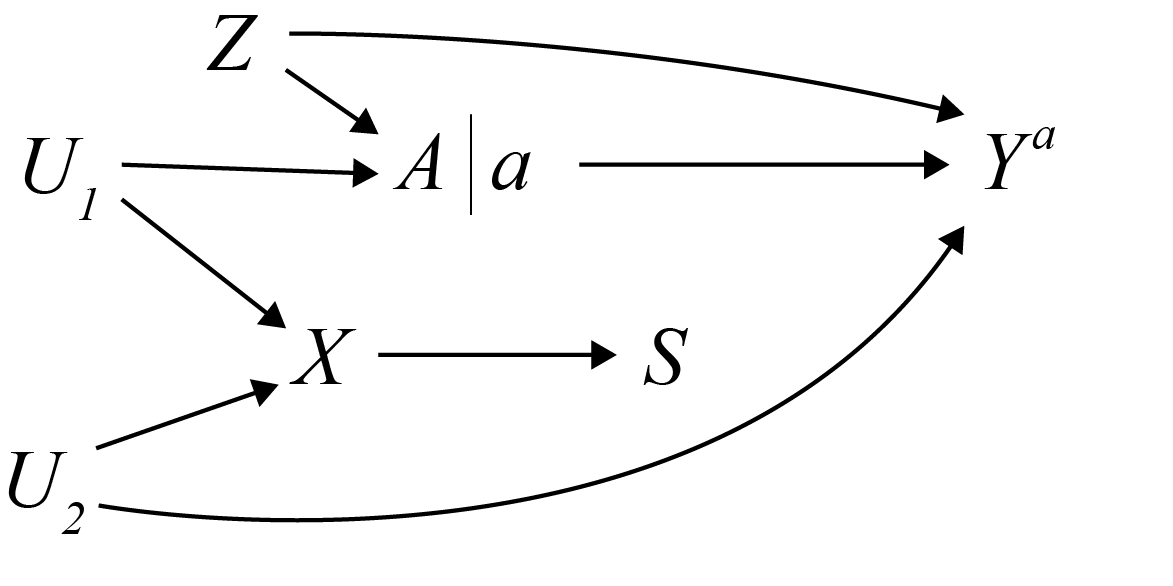}
	\label{Figure3}
\end{figure}

Now, consider the joint occurrence of confounding and selection bias described in Zivich et al. and shown in Figure \ref{Figure3} \cite{Zivich2022On}. Note that we recycle our notation (i.e., variable definitions do not carry over from Case 1). Let $A$ denote selective serotonin reuptake inhibitor use and $Y$ indicate incident lung cancer in the following 10 years. Again, we are interested in estimated the average causal effect, $\psi$. Let $Z$ denote smoking status, $X$ denote cardiovascular risk score, $U_1$ denote depression, $U_2$ denote occupational exposures, and $S$ denote if a person completed follow-up. Both $U_1$ and $U_2$ were unmeasured. As indicated by Figure \ref{Figure3}, $Z$ is a confounding variable, and selection bias was related to $X$. Here, there is no superset of variables that addresses both biases. Adjusting for $\{X,Z\}$ opens a backdoor path from $A$ to $Y^a$, and only adjusting for $\{X\}$ or $\{Z\}$ does not address both confounding and selection bias. Instead, the parameter of interest can be written as
\begin{equation}
	\mu^a = E \left\{ E \left[ E(Y \mid A=a, Z, X, S=1) \mid A=a, Z \right] \right\}
	\label{ID-R3}
\end{equation}
with a proof provided in Appendix 3.

\subsection{Building an estimator}

The identification result in \ref{ID-R3} indicates how we an build an iterated g-computation estimator. Starting with the innermost expectation, a model for $E(Y \mid A,Z,X,S=1)$ is estimated. Using that model, we generate pseudo-outcomes under $a$ for all observations, $\hat{Y}^a$. Moving to the next expectation, we now fit a model for the pseudo-outcomes given the observed $A$ and $Z$, $E[\hat{Y}^a \mid A,Z]$. Note that the identification results suggest that a model for this value could be separately estimated by levels of $A$ (or equivalently by including interaction terms with $A$ for all covariates in $Z$). However, this need not be the case for the estimator if we assume the corresponding parametric model is correctly specified. In practice, fitting stratified models (or including all the interactions) will likely be preferred due to the weaker modeling assumptions being imposed. Regressing pseudo-outcomes is an approach also used by the iterated conditional expectation g-computation algorithm for time-varying confounding \cite{Zivich2024Empirical}, a modification of g-computation to address measurement error with partial colliders \cite{RossLeveraging}, estimating the conditional average causal effect \cite{Laan2015Targeted, Zivich2024Synthesis}, and estimating the parameters of a marginal structural model \cite{Zivich2025Estimating, Snowden2011Implementation}. This second model is then used to simulate another set of pseudo-outcomes setting $A$ to $a$, which we denote as $\tilde{Y}^a$. The mean of $\tilde{Y}^a$ is then our estimate of $\mu^a$. 

Again, these steps can be translated into a set of estimating functions. The corresponding estimating functions are
\begin{equation*}
	g_3(Z_i; \theta) = 
	\begin{bmatrix}
		S_i \left[ Y_i - m(\mathbb{W}_i; \beta)  \right] \mathbb{W}_i^T \\
		\left[ \hat{Y}_i^1 - m(\mathbb{V}_i; \gamma_1)  \right] \mathbb{V}_i^T \\
		\tilde{Y}_i^1 - \mu^1 \\
		\left[ \hat{Y}_i^0 - m(\mathbb{V}_i; \gamma_0)  \right] \mathbb{V}_i^T \\
		\tilde{Y}_i^0 - \mu^1 \\
		(\mu^1 - \mu^0) - \psi
	\end{bmatrix}
\end{equation*}
where $\theta = (\beta, \gamma_1, \mu^1, \gamma_0, \mu^0, \psi)$, $m(\mathbb{W}; \beta) = E[Y \mid A,Z,X; \beta]$ with the design matrix $\mathbb{W}$, and $m(\mathbb{V}; \gamma_a) = E[\hat{Y}^a \mid A,Z]$ with the design matrix $\mathbb{V}$. The first estimating function is again the outcome model restricted to those who completed follow-up. The second estimating function is now a regression of $\hat{Y}^1$ given the observed $A,Z$ for all observations. The third is the mean of the updated pseudo-outcomes with $a:=1$. The fourth and fifth estimating functions repeat the process for $a:=0$. The final estimating function is for the average causal effect.

Now, we argue that our proposed estimator is consistent and asymptotically normal with a more formal proof in Appendix 3. Note that the outcome model is unbiased following from maximum likelihood theory. Similarly, unbiasedness of $E\left\{ \left[ \hat{Y}_i^a - m(\mathbb{V}_i; \gamma_a)  \right] \mathbb{V}_i^T \right\}$ follows from quasi-maximum likelihood \cite{Papke1996Econometric, Godambe1987Quasi}. The unbiasedness of the means follows from the identification assumptions and correct model specification. Finally, the estimating equation for the average causal effect follows from the same arguments as before.

\subsection{Simulation}

Again, we verify our proposed estimator using a simulation study. A modified version of the data generating mechanism from Ross et al. was created (see Appendix 3 for details) \cite{RossLeveraging}. Here, we compared a na\"ive complete-case analysis, standard g-computation adjusting for only $Z$, standard g-computation adjusting for only $X$, standard g-computation adjusting for $\{X,Z\}$, our proposed iterated g-computation algorithm, and an IPW estimator for reference (Appendix 3) \cite{Zivich2022On}. The same set of evaluation metrics from case 1 were considered for a simulation of $n:=1000$ with 5000 repetitions. Simulations were conducted in Python 3.9.4 with the same set of packages. Again, we also replicate the results for a single simulated data set in R.

The iterated g-computation and IPW estimators had negligible bias and appropriate CI coverage (Table \ref{Table2}), agreeing with the theoretical expectations. All other approaches had bias and below nominal CI coverage. The iterated g-computation estimator was the best performing estimator in terms of the RMSE.

\begin{table}[H]
	\caption{Simulation results for confounding and selection bias lacking a superset}
	\centering		
	\begin{tabular}{lccccc}
		\hline
		& Bias  & ESE & RMSE & SER  & Coverage \\ \cline{2-5} 
		Complete-case analysis (na\"ive) & 0.037 & 0.034 & 0.050 & 1.01 & 0\%      \\
		Standard g-computation, $\{Z\}$                 & 0.025 & 0.033 & 0.041 & 1.01 & 90\%     \\
		Standard g-computation, $\{X\}$                 & 0.042 & 0.039 & 0.058 & 1.00 & 83\%     \\
		Standard g-computation, $\{X,Z\}$               & 0.029 & 0.038 & 0.047 & 1.00 & 90\%     \\
		Iterated g-computation                          & 0.004 & 0.036 & 0.036 & 1.00 & 95\%     \\
		Inverse probability weighting                   & 0.000 & 0.038 & 0.038 & 1.00 & 95\%     \\ \hline
	\end{tabular}
	\floatfoot{ESE: empirical standard error, RMSE: root mean squared error, SER: standard error ratio, CI: confidence interval. \\
	Simulation results were based on a sample size of 1000 observations repeated for 5000 iterations.}
	\label{Table2}
\end{table}

\section{Discussion}

Here, we provided two case studies in the context of selection bias on how to develop novel g-computation estimators. These examples illustrate how epidemiologists can translate causal diagrams or identification results into estimators and review some of their properties both theoretically and empirically. The basic principles outlined here can be extended to build g-computation estimators for more complex causal structures or with additional sources of bias. Furthermore, we found that these basic principles are particularly useful in the scenarios with selection bias, since (1) nonrandom selection (sampling) is often accompanied by missing data on some important variables and then leads to different types of selection bias \cite{lu2024revisiting}, and (2) researchers tend to rely on na\"ive complete-case analyses or standard approaches to adjusting for covariates that are related to selection bias, which can result in substantial bias for some nonrandom selection mechanisms.

To develop our estimators and prove some of their theoretical properties, we relied on M-estimation theory. While M-estimators are not required, we find the additional step of translating from discrete steps into estimating functions to be worth the additional effort (e.g., simplified asymptotic proofs, variance estimation). However, there are important limitations to this framework one should be aware of. First, standard M-estimation theory assumes that $\theta$ is of finite-dimension \cite{Boos2013M}. As such, nonparametric estimators (e.g., Kaplan-Meier) are not immediately ammenable to this framework. Other theoretical justifications \cite{2008Z, Breslow2015Z}, or structural assumptions are needed \cite{zivich_estimating_2025}. Further, M-estimation assumes that there is a single solution to the estimating equations, which precludes estimators with objective functions with multiple maxima (e.g., neural networks). A third issue is that there are certain restrictions on the structure of the estimating functions, which are reviewed in-depth elsewhere \cite{Boos2013M}. Notables examples that prevent use of the standard empirical sandwich variance estimator are the Cox proportional hazards model and the L1 penalty (i.e., LASSO) \cite{Lin1989Robust, fu_penalized_1998, fu_penalized_2003}.  Fourth, all relevant estimation steps must be included in the stacked estimating functions for valid inference with the empirical sandwich variance estimator. Importantly, this included model or variable selection steps that are not based on subject matter knowledge. While the L1 penalty presents issues for inference, the L2 penalty or smoothed versions of the L1 penalty provide ways forward \cite{fu_penalized_1998, fu_penalized_2003, haselimashhadi_unified_2019}. Despite these important limitations, M-estimation still covers a broad range of applications in epidemiology.

There are also some challenges to developing estimators that epidemiologists should be aware of. First, identification does not necessarily imply that a parameter is estimable with data, even as the number of observations grows to infinity \cite{Maclaren2020What, Aronow2021Nonparametric}. This problem arose in the derivation of our estimators, where we also assumed that our parametric model for the outcome process was correctly specified. Second, the proofs of unbiased estimating equations were straightforward for both our estimators. This was mainly due to each estimating function being based on regression models or the mean. However, these proofs may not always be trivial. Therefore, close collaborations with statisticians, biostatisticians, and others experienced in statistical theory are still recommended.

There are several directions in which this work could be further developed. First, one might consider how models for different processes can be combined to construct augmented inverse probability weighting (AIPW) or targeted maximum likelihood estimators \cite{schuler_targeted_2017, funk_doubly_2011}, which may offer robustness to different modeling assumptions. One approach to deriving an AIPW estimator is to derive the influence function from the identification results, which can be done using several different methods \cite{hines_demystifying_2022, renson_pulling_2025}. Given the connection between estimating function and influence functions \cite{cole_five_2025}, that AIPW estimator could then be expressed with estimating functions to possibly allow for multiply robust point and variance estimation with parametric models \cite{shook-sa_double_2025}. There also might be interest in using flexible machine learning algorithms instead of parametric regression models. While the use of machine learning for estimation presents certain challenges, these can often be addressed through the use of AIPW estimators \cite{Daniel2018Double, chernozhukov_doubledebiased_2018, zivich_machine_2021, zivich_machine_2022}. Again, it is helpful to turn to influence functions here. Influence functions provide ways to study additional asymptotic properties of estimators, including convergence rates with flexible machine learning algorithms \cite{renson_pulling_2025}. While the influence function approach can provide additional results, g-computation estimators built with estimating functions can be directly developed from identification results and thus are a useful first step. Next, our examples did not include the setting where outcomes influence selection (e.g., case-control studies). In these settings, identifying average cusal effects in the entire sample might be challenging. Ultimately, the recoverability of causal effects in the presence of selection bias will depend on the underlying causal structure, the information available, and the parameter of interest. While the proposed g-computation estimators may no longer apply, the underlying ideas of building up estimators from identification results will. Finally, the aim of our simulations was to verify the derivations and showcase some general finite sample performance, so selection bias in our mechanisms was quite substantial. Exploring the properties of all the competing estimators under different selection mechanisms with varying relationship strengths and sample sizes or mimicing particular use-cases in future work would be informative to guide their practical use \cite{Vaughan2009use}.

To conclude, the translation from causal diagrams to estimators is an essential step in epidemiology. For some causal structures, like the reviewed selection bias mechanisms, well-known estimation procedures may need to be modified. Here, we illustrated the process of translating the identification results of a causal diagram into a g-computation estimator using estimating equations. This framework provides a powerful set of tools for epidemiologists to address many practical problems.

\section*{Acknowledgments}

Research reported in this publication was supported by National Institute of Allergy and Infectious Diseases and the National Institute on Drug Abuse of the National Institutes of Health under award numbers K01AI177102 (PNZ), and K99DA057487 (HL). The content is solely the responsibility of the authors and does not necessarily represent the official views of the National Institutes of Health. Python and R code to replicate the analysis and simulations is available at \url{https://github.com/pzivich/publications-code}

{\small
\bibliography{references}{}
\bibliographystyle{ieeetr}}

\newpage 

\section*{Appendix}

\subsection*{Appendix 1: Standard G-computation}

\subsubsection*{A1.1: Identification Proof}

Here, we prove the identification result provided in \ref{ID-R1}. Note that we assume conditional treatment exchangeability ($Y^a \amalg A \mid X$) with positivity ($\Pr(A=a \mid X=x) > 0$ for all $a \in \{0,1\}$ and $x \in \mathcal{X}$ where $\mathcal{X}$ is the support of $X$), conditional selection exchangeability ($Y^a \amalg S \mid A,X$) with positivity ($\Pr(S=1 \mid A=a,X=x) > 0$ for all $a \in \{0,1\}$ and $x \in \mathcal{X}$), and causal consistency ($Y_i^{A_i} = Y_i$). Therefore, 
\begin{equation}
	\begin{aligned}
		\mu^a = E[Y^a] & = E[E(Y^a \mid X)] \\
		& = E[E(Y^a \mid A=a, X)] \\
		& = E[E(Y^a \mid A=a, X, S=1)] \\
		& = E[E(Y \mid A=a, X, S=1)] \\
	\end{aligned}
\end{equation}
following from the definition of $\mu^a$, the law of iterated expectation, conditional treatment exchangeability with positivity, conditional selection exchangeability with positivity, and causal consistency, respectively.

\subsubsection*{A1.2: Consistency and Asymptotic Normality}

Recall that the estimating function for standard g-computation are
\begin{equation*}
	g_1(Z_i; \theta) = 
	\begin{bmatrix}
		S_i \left[ Y_i - m(\mathbb{X}_i; \beta)  \right] \mathbb{X}_i^T \\
		\hat{Y}_i^1 - \mu^1 \\
		\hat{Y}_i^0 - \mu^0 \\
		(\mu^1 - \mu^0) - \psi
	\end{bmatrix}
\end{equation*}
To begin, consider the first estimating function for a generic element $v \in \mathbb{X}$ without a loss of generality. Note that
\begin{equation*}
	\begin{aligned}
		E \left\{ S \left[ Y - m(\mathbb{X}; \beta) \right] v \right\} & = E \left\{ \left[ Y - m(\mathbb{X}; \beta) \right] v \mid S=1 \right\} \Pr(S=1) \\
		& = E \left\{ Yv - m(\mathbb{X}; \beta)v  \mid S=1 \right\} \Pr(S=1) \\
		& = \left\{ E\left[ Yv \mid S=1\right] - E\left[ m(\mathbb{X}; \beta)v  \mid S=1\right]  \right\} \Pr(S=1) \\
		& = \left\{ E\left[ E(Yv \mid A,X,S=1) \mid S=1\right] - E\left[ m(\mathbb{X}; \beta)v  \mid S=1\right]  \right\} \Pr(S=1) \\
		& = \left\{ E\left[ v E(Y \mid A,X,S=1) \mid S=1\right] - E\left[ m(\mathbb{X}; \beta)v  \mid S=1\right]  \right\} \Pr(S=1) \\
		& = E\left[ v \{E(Y \mid A,X,S=1) - m(\mathbb{X})\} \mid S=1 \right] \Pr(S=1) \\
		& = E\left[ v \{E(Y \mid A,X,S=1) - E(Y \mid A,X,S=1; \beta)\} \mid S=1 \right] \Pr(S=1) \\
		& = E\left[ v \{E(Y \mid A,X,S=1) - E(Y \mid A,X,S=1)\} \mid S=1 \right] \Pr(S=1) \\
		& = 0
	\end{aligned}
\end{equation*} 
following from the law of total expectation, distribution of $v$, sum of expectations, law of iterated expectations, $v$ must be either $A,X$, or some transformation of $A,X$, sum of expectations, definition of $m(\mathbb{X}; \beta)$, correct model specification, and identity. As this was proven for a generic element in the design matrix, this proof of unbiasedness suffices to show the estimating equation is unbiased for all elements in $\mathbb{X}$ Additionally, if one adopts a perspective of a regression as an orthogonal projection, then correct model specification is not necessary as an assumption here. As the following proofs require correct model specification, we assume so here as well to simplify exposition. Now consider the second and third estimating functions. For ease, we prove
\begin{equation*}
	\begin{aligned}
		E[\hat{Y}^a - \mu^a] & = E[\hat{Y}^a] - \mu^a \\
		& = E[E(Y \mid A=a,X,S=1)] - \mu^a \\
		& = E[Y^a] - \mu^a \\
		& = E[Y^a] - E[Y^a] = 0 \\
	\end{aligned}
\end{equation*}
for $a \in \{0,1\}$. Here, the first step follows from the expectation of a constant, the second follows from the definition of $\hat{Y}^a$ and correct model specification, the third step follows from the identification proof, and the final step follows from the definition of $\mu^a$. Finally, it immediately follows that $E[(\mu^1 - \mu^0) - \psi] = 0$ from the definitions of $\psi$ and $\mu^a$.

Having show each of the corresponding estimating equations to be unbiased at $\theta$, it follows under suitable regularity conditions that $\sqrt{n}(\hat{\theta} - \theta) \rightarrow^d N(0, V(\theta))$ (i.e., the difference between our estimator and the truth converges to a normal distribution with variance $V(\theta)$ at a rate of $\sqrt{n}$), where $V(\theta) = B^{-1}(\theta) M(\theta) [B^{-1}]^T $ is the sandwich variance with $B(\theta) = E[-\partial g(\theta) / \partial \theta]$ is the expectation of the gradient of the estimating functions and $M(\theta) = E[g(\theta) g(\theta)^T]$ is the expectation of the outer product of the estimating functions \cite{Boos2013M}.

\subsection*{Appendix 2: G-computation for Case 1}

\subsubsection*{A2.1: Identification Proof}

Here, we prove the identification result provided in \ref{ID-R2}. As implied by Figure \ref{Figure1}, we now assume marginal treatment exchangeability ($Y^a \amalg A$) with positivity ($\Pr(A=a) > 0$ for all $a \in \{0,1\}$). For conditional selection exchangeability, we assume that $Y^a \amalg S^a \mid A=a, X^a$, as implied by the SWIG. This assumption comes paired with the probability assumption that $\Pr(S=1 \mid A=a, X^a = x) > 0$ for all $a \in \{0,1\}$ and $x \in \mathcal{X}^a$. Finally, we assume causal consistency (i.e., $X^A = X$, $S^A = S$, $Y^A = Y$). Therefore,
\begin{equation*}
	\begin{aligned}
		\mu^a = E[Y^a] & = E[Y^a \mid A=a ] \\
		& = E[E(Y^a \mid A=a, X^a) \mid A=a ] \\
		& = E[E(Y^a \mid A=a, X^a, S^a = 1) \mid A=a ] \\
		& = E[E(Y \mid A=a, X, S = 1) \mid A=a ] \\
	\end{aligned}
\end{equation*}
following from marginal exchangeability with positivity, law of iterated expectations, conditional selection exchangeability with positivity, and causal consistency, respectively.

\subsubsection*{A2.2: Consistency and Asymptotic Normality}

Recall that the proposed g-computation estimating functions are
\begin{equation*}
	g_2(Z_i; \theta) = 
	\begin{bmatrix}
		S_i \left[ Y_i - m(\mathbb{X}_i; \beta)  \right] \mathbb{X}_i^T \\
		A_i \left(\hat{Y}_i^1 - \mu^1\right) \\
		(1 - A_i) \left(\hat{Y}_i^0 - \mu^0\right) \\
		(\mu^1 - \mu^0) - \psi
	\end{bmatrix}
\end{equation*}
It follows that the first estimating function is unbiased following the same steps of the proof provided in A1.2. Now consider the second estimating function for $a \in \{0,1\}$,
\begin{equation*}
	\begin{aligned}
		E\left[ I(A=a) (\hat{Y}^a - \mu^a) \right] & = E\left[ \hat{Y}^a \mid A=a \right] \Pr(A=a) - E\left[ \mu^a \mid A=a \right] \Pr(A=a) \\
		& = E\left[ \hat{Y}^a \mid A=a \right] \Pr(A=a) - \mu^a \Pr(A=a) \\
		& = E\left[ E(Y \mid A=a, X) \mid A=a \right] \Pr(A=a) - \mu^a \Pr(A=a) \\
		& = E\left[ Y^a \right] \Pr(A=a) - \mu^a \Pr(A=a) \\
		& = E\left[ Y^a \right] \Pr(A=a) - E[Y^a] \Pr(A=a) = 0 \\
	\end{aligned}
\end{equation*}
following from the law of total exchangeability, $\mu^a$ being a constant, definition of $\hat{Y}^a$ under correct model specification, the identification proof, and the definition of $\mu^a$, respectively. Therefore, both the second and third estimating equations are unbiased. Again, the final estimating equation is unbiased following the same proof as the one provided in Appendix 1.2. Therefore, it follows that $\sqrt{n} (\hat{\theta} - \theta) \rightarrow^d N(0, V(\theta))$.

\subsubsection*{A2.3: Data Generating Mechanism}

The following data generating mechanism was used:
\begin{equation*}
	\begin{aligned}
		A & \sim \text{Beroulli}(0.5) \\
		U & \sim \text{Normal}(\mu=0, \sigma=1) \\
		X & \sim \text{Normal}(\mu=-1+2A+U, \sigma=1) \\
		S & \sim \text{Bernoulli}(\text{expit}(2 - 1X)) \\
		Y & \sim 
		\begin{cases}
			\text{Bernoulli}(\text{expit}(0.5 + 0.75U - 1A)) & \text{ if } S=1 \\
			-9999 & \text{ if } S=0 \\
		\end{cases}
	\end{aligned}
\end{equation*}
The true value of $\psi$ (as approximated by simulation 10 million observations with $S:=1$ for all observations) was $-0.219$. In the observed data, approximately 20\% of participants had missing values of $Y$.

\subsubsection*{A2.4: Inverse Probability Weighting Estimator}

The corresponding IPW expression for the identification results in A2.1 is
\begin{equation*}
	\mu^a = E \left[ Y \times I(A=a) \times \frac{S}{\Pr(S=1 \mid X)} \right]
\end{equation*}
and the corresponding estimating functions for the Hajek IPW are
\begin{equation*}
	\begin{bmatrix}
		\left\{ S_i - \text{expit}(\mathbb{S}_i \hat{\alpha}^T) \right\} \mathbb{S}_i^T \\
		\frac{S_i}{\text{expit}(\mathbb{S}_i \hat{\alpha}^T)} A_i (Y_i - \hat{\mu}^1) \\ 
		\frac{S_i}{\text{expit}(\mathbb{S}_i \hat{\alpha}^T)} \left\{1 - A_i\right\} (Y_i - \hat{\mu}^0) \\ 
		(\hat{\mu}^1 - \hat{\mu}^0) - \hat{\psi}
	\end{bmatrix}
\end{equation*}
where $\text{expit}(\mathbb{S}_i \hat{\alpha}^T)$ is a logistic regression model for the probability of $S$ given $X$ (i.e., $\mathbb{S}$ is a design matrix that includes $X$, $\text{expit}(x) = \{1 + \exp(-x)\}^{-1}$). Note that we do not provide a proof of the asymptotic properties of this estimators as it is beyond the scope of this paper.

\subsection*{Appendix 3: G-computation for Case 2}

\subsubsection*{A3.1: Identification Proof}

As in the main paper, notation is recycled from the prior case study. Here, we prove the identification result provided in \ref{ID-R3}. As implied by Figure \ref{Figure2}, we have the following exchangeability assumptions (with their corresponding positivity assumptions): $Y^a \amalg A \mid Z$ and $Y^a \amalg A,Z,X$. Again, we assume causal consistency. Therefore,
\begin{equation*}
	\begin{aligned}
		\mu^a = E\left[ Y^a \right] & = E\left[ E(Y^a \mid Z) \right] \\
		& = E\left[ E(Y^a \mid A=a, Z) \right] \\
		& = E\left\{ E\left[ E(Y^a \mid A=a, Z, X) \mid A=a, Z \right] \right\}  \\
		& = E\left\{ E\left[ E(Y^a \mid A=a, Z, X, S=1) \mid A=a, Z \right] \right\}  \\
		& = E\left\{ E\left[ E(Y \mid A=a, Z, X, S=1) \mid A=a, Z \right] \right\}  \\
	\end{aligned}
\end{equation*}
following from the law of iterated expectations, conditional treatment exchangeability with positivity, the law of iterated expectations, conditional selection exchangeability with positivity, and causal consistency, respectively.

\subsubsection*{A3.2: Consistency and Asymptotic Normality}

Again, recall that the stacked estimating functions are
\begin{equation*}
	g_3(Z_i; \theta) = 
	\begin{bmatrix}
		S_i \left[ Y_i - m(\mathbb{W}_i; \beta)  \right] \mathbb{W}_i^T \\
		\left[ \hat{Y}_i^1 - m(\mathbb{V}_i; \gamma_1)  \right] \mathbb{V}_i^T \\
		\tilde{Y}_i^1 - \mu^1 \\
		\left[ \hat{Y}_i^0 - m(\mathbb{V}_i; \gamma_0)  \right] \mathbb{V}_i^T \\
		\tilde{Y}_i^0 - \mu^1 \\
		(\mu^1 - \mu^0) - \psi
	\end{bmatrix}
\end{equation*}
A similar proof to the one in Appendix 1.2 can be used to show the first estimating equation is unbiased. Now consider the second and fourth estimating functions, consider the generic element $v \in \mathbb{V}$ for $a \in \{0,1\}$
\begin{equation*}
	\begin{aligned}
		E \left\{ [\hat{Y}^a - m(\mathbb{V}; \gamma_a)] v \right\} & = E[\hat{Y}^a v] - E[m(\mathbb{V}; \gamma_a) v] \\
		& = E[\hat{Y}^a v] - E[E(\hat{Y}^a \mid A,Z) v] \\
		& = E[E(\hat{Y}^a v \mid A,Z)] - E[E(\hat{Y}^a \mid A,Z) v] \\
		& = E[E(\hat{Y}^a \mid A,Z) v] - E[E(\hat{Y}^a \mid A,Z) v] = 0\\
	\end{aligned}
\end{equation*}
following from the distributive property and sum of expectations, the definition of $m(\mathbb{V}; \gamma_a)$, the law of iterated expectations, and $v$ being some combination of $A,Z$. Again,
\begin{equation*}
	\begin{aligned}
		E[\tilde{Y}^a - \mu^a] & = E[\tilde{Y}^a] - \mu^a \\
		& = E\{E[E(Y \mid A=a, Z,X,S=1) \mid A=a, Z]\} - E[Y^a] \\
		& = E[Y^a] - E[Y^a] = 0 \\
	\end{aligned}
\end{equation*}
for $a \in \{0,1\}$. Again, it straightforwardly follows that the estimating equation for the average causal effect is unbiased following the definition of the parameters. Therefore, it follows that $\sqrt{n} (\hat{\theta} - \theta) \rightarrow^d N(0, V(\theta))$.

\subsubsection*{A3.3: Data Generating Mechanism}

The following data generating mechanism was used
\begin{equation*}
	\begin{aligned}
		U_1 & \sim \text{Beroulli}(0.5) \\
		U_2 & \sim \text{Beroulli}(0.5) \\
		Z & \sim \text{Beroulli}(0.5) \\
		A & \sim \text{Bernoulli}(\text{expit}(-2.3 + \log(2) Z + \log(4) U_1)) \\
		X & \sim \text{Normal}(\mu=4U_1 - 4 U_2, \sigma=1) \\
		S & \sim \text{Bernoulli}(\text{expit}(0.25X)) \\
		Y & \sim 
		\begin{cases}
			\text{Bernoulli}(\text{expit}(-2 -2A \log(2) Z + \log(2) AZ + \log(4) U_2)) & \text{ if } S=1 \\
			-9999 & \text{ if } S=0 \\
		\end{cases}
	\end{aligned}
\end{equation*}
The true value of $\psi$ (as approximated by simulation 10 million observations with $S:=1$ for all observations) was $-0.205$. In the observed data, approximately 50\% of participants had missing values of $Y$.

\subsubsection*{A2.4: Inverse Probability Weighting Estimator}

The corresponding IPW expression for the identification results in A3.1 is
\begin{equation*}
	\mu^a = E \left[ Y \times \frac{I(A=a)}{\Pr(A=a \mid Z)} \times \frac{S}{\Pr(S=1 \mid X)} \right]
\end{equation*}
and the corresponding estimating functions for the Hajek IPW are
\begin{equation*}
	\begin{bmatrix}
		\left\{ S_i - \text{expit}(\mathbb{S}_i \hat{\alpha}^T) \right\} \mathbb{S}_i^T \\
		\left\{ A_i - \text{expit}(\mathbb{A}_i \hat{\delta}^T) \right\} \mathbb{A}_i^T \\

		\frac{S_i}{\text{expit}(\mathbb{S}_i \hat{\alpha}^T)} \times \frac{A_i}{\text{expit}(\mathbb{A}_i \hat{\delta}^T)} \times (Y_i - \hat{\mu}^1) \\ 
		\frac{S_i}{\text{expit}(\mathbb{S}_i \hat{\alpha}^T)} \times \frac{1-A_i}{1-\text{expit}(\mathbb{A}_i \hat{\delta}^T)} \times (Y_i - \hat{\mu}^0) \\ 
		(\hat{\mu}^1 - \hat{\mu}^0) - \hat{\psi}
	\end{bmatrix}
\end{equation*}
where $\text{expit}(\mathbb{S}_i \hat{\alpha}^T)$ is a logistic regression model for the probability of $S$ given $X$ (i.e., $\mathbb{S}$ is a design matrix that includes $X$) and $\text{expit}(\mathbb{A}_i \hat{\delta}^T)$ is a logistic regression model for the probability of $A$ given $Z$ (i.e., $\mathbb{A}$ is a design matrix that includes $Z$). Note that we do not provide a proof of the asymptotic properties of this estimators as it is beyond the scope of this paper.

\end{document}